\documentclass[conference, 12pt]{IEEEtran}
\IEEEoverridecommandlockouts
\usepackage{cite}
\usepackage{amsmath,amssymb,amsfonts}
\usepackage{algorithmicx,algorithm}
\usepackage{graphicx}
\usepackage{textcomp}
\usepackage{xcolor}
\usepackage{multirow}
\usepackage[colorlinks=true, allcolors=blue]{hyperref}

\def\BibTeX{{\rm B\kern-.05em{\sc i\kern-.025em b}\kern-.08em
    T\kern-.1667em\lower.7ex\hbox{E}\kern-.125emX}}
\begin{document}
\pagestyle{plain}
\title{Cyber-Physical Taint Analysis in Multi-stage Manufacturing Systems (MMS): A Case Study}

\author{\IEEEauthorblockN{Tao Liu\IEEEauthorrefmark{1}, Bowen Yang\IEEEauthorrefmark{2}, Qi Li\IEEEauthorrefmark{2}, Jin Ye\IEEEauthorrefmark{2}, Wenzhan Song\IEEEauthorrefmark{2}, Peng Liu\IEEEauthorrefmark{1}}
\IEEEauthorblockA{{\IEEEauthorrefmark{1}College of Information Sciences and Technology, Pennsylvania State University, State College, PA, USA} 
\\\{tul459, pxl20\}@psu.edu
}

\IEEEauthorblockA{{\IEEEauthorrefmark{2} School of Electrical and Computer Engineering, University of Georgia, Athens, GA, USA} 
\\{\{bowen.yang, ql61608, jin.ye, wsong\}@uga.edu}
}

}
\maketitle
\begin{abstract}
Information flows are intrinsic properties of an multi-stage manufacturing systems (MMS). In computer security, a basic information flow tracking technique is dynamic taint analysis (DTA).  DTA tracks taint propagation from one data variable (e.g., a buffer holding a HTTP request) to another. Taint propagation paths are typically determined by data flows and implicit
flows in a computer program. And the union of all the taint propagation paths forms a taint graph. It is clear that taints graphs could
significantly enhance intrusion diagnosis. However, the existing DTA techniques cannot be directly used in an MMS, and a main reason is as follows: Without manufacturing-specific taint propagation rules, DTA cannot be implemented. 
In this work, we conduct a case study which (a) extends the existing DTA method with manufacturing-specific taint propagation rules, and (b) applies the extended method to perform preliminary intrusion diagnosis with a small-scale test-bed.  
\end{abstract}


\section{Introduction}
\label{1.Intro}

Multi-stage Manufacturing Systems (MMS) are playing an important role in the manufacturing industry. 
In the manufacturing sector, automation, interconnection and electrification are increasingly adopted for better efficiency and cost-reduction. With the increase in software-based networking, monitoring, and control of manufacturing assets across networks, the risk of cyber-attacks also grows. Cybersecurity challenges in the manufacturing sector are serious and have significant impacts on the competitiveness of the national economy and defense. More than 400 manufacturers were targeted every day in 2016 and almost half of them were US manufacturers, resulting in more than 3 billion in losses \cite{mahoney2017cybersecurity}. The variety and scale of cyber-sabotage on manufacturing systems have been growing in recent years and received the increasing attention due to its catastrophic consequences to the economy, people and environment. Manufacturing systems are subject to a wide range of passive and active cyber threats ranging from data and intellectual property (IP) theft to cyber-sabotage that alter the quality of the product or destroy manufacturing machines, reducing manufacturing productivity, and increasing costs \cite{mahoney2017cybersecurity, tuptuk2018security, langner2011stuxnet}.

Most modern manufacturing systems are Multistage Manufacturing Systems (MMS) \cite{shi2006stream}, which consists of multiple components, stations, or stages to finish the final product. MMS is an engineering designed system, with the specific designated control sequence and features to be fabricated. Thus, the cyber and physical signals would have deterministic sequences and patterns in normal situations. Also, the cyber signals and physical signals are interrelated in a manufacturing system, which gives opportunities for data fusion to jointly analyze cyber and physical signals together. The quality of the final product is determined by complex interactions among multiple stages - the quality characteristics at one stage are not only influenced by local variations at that stage, but also by variations propagated from upstream stages. Thus, a cyber-threat may intrude a controller/machine in one stage and impact other stages through the cyber networks or physical networks, which makes the cyber-threat detection and diagnosis in MMS complex, challenging and interesting.


In this work, we seek to conduct a case study on the usefulness of cyber-physical information flow analysis in performing intrusion diagnosis in an MMS.  Information flows are intrinsic properties of an MMS. In computer security, a basic information flow tracking technique is dynamic taint analysis (DTA).  DTA tracks taint propagation from one data variable (e.g., a buffer holding a HTTP request) to another. Taint propagation paths are typically determined by data flows and implicit
flows in a computer program. And the union of all the taint propagation paths forms a taint graph. It is clear
that effective intrusion monitoring can be done via information flow tracking, and that taints graphs could
significantly enhance intrusion diagnosis. However, the existing DTA techniques cannot be directly used in an MMS, and a main reason is as follows:  Without manufacturing-specific taint propagation rules, DTA cannot be implemented. For example, when a motor drive receives a distorted control signal from a compromised PLC, although the signal is a variable of the control logic program, the result (of the motor motion) is no longer a variable in any program. Restricted to cyberspace, no existing taint propagation rule can be used to propagate the taint onto the motion result. 

\begin{figure*}[htbp]
\centering
\centerline{\includegraphics[width=0.8\textwidth]{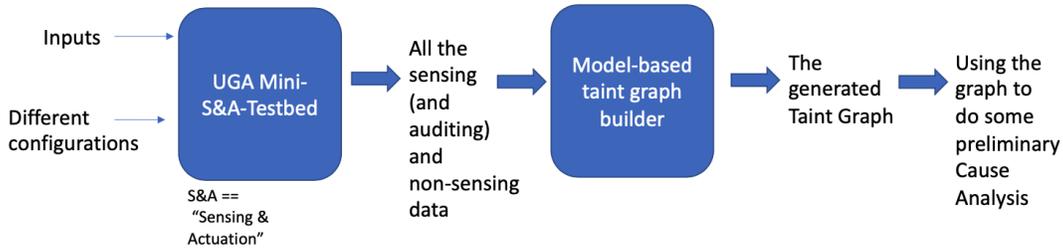}}
\caption{Workflow Overview}
\label{emu}
\end{figure*}

In this paper, a case study is conducted to test the effectiveness of a set of manufacturing-specific taint propagation rules proposed by us. In particular, the case study consists of the following activities and findings:

\begin{itemize}
\item We have developed a preliminary generic cyber-physical information flow model for an MMS. The model identifies several types of manufacturing-specific security-related information flows. The model also identifies a set of manufacturing-specific taint propagation rules. 

\item A scaled-down testbed has been built to collect MMS-specific event data.  

\item We have used the specific event data collected from the specific testbed to customize the generic cyber-physical information flow model. The customized model can be used to explain the information flow semantics reflected in the specific event data. 

\item Using the customized set of manufacturing-specific taint propagation rules, we have developed a software tool to construct a taint graph directly based on the event data collected from the testbed. 

\item Through answering several specific intrusion diagnosis questions, including (a) ``if control signal X is compromised, would Y be part of the damage?" and (b) ``if damage Y is observed, would X be a causing factor?", we have been evaluating the soundness and effectiveness of the proposed cyber-physical information flow model for an MMS and the manufacturing-specific taint propagation rules. 
\end{itemize} 

To the best of our knowledge, this case study is the first one doing Cyber-Physical Taint Analysis (for intrusion diagnosis) in Multi-stage Manufacturing Systems.   

The remaining of this paper is organized as follows. Section \ref{Back} provides preliminary introduction of Multistage Manufacturing System, threat model, and the existing works. Section \ref{Over} brings out the both challenges and insights of how to extend the existing DTA method in an Multi-stage Manufacturing System (MMS). Following the workflow shown in Figure \ref{emu}, Section \ref{UGA} introduces the ``Mini-S\&A-Testbed" used in this work and how the data is collected. Our proposed information flow model is demonstrated in Section \ref{Infor}. Section \ref{Taint} illustrates how the taint graph is generated based on both information flow model and collected data set. Section \ref{Case} provides two representative case studies for cause analysis. Section \ref{Con} concludes this work.

\section{Background}
\label{Back}

\subsection{Preliminary Introduction of Multistage Manufacturing System}
Multistage Manufacturing System (MMS) refers to a manufacturing system consisting of multiple components, stations, or stages required to finish the final product. Almost all modern manufacturing processes (e.g., assembly, machining, semiconductor fabrication, pharmaceutical manufacturing) fit this category. In this {\color{red}report}, a multi-stage hot rolling mill automation system testbed will be used for case studies and experiments. It consists of digital motor drives, Programmable Logic Controllers (PLC), control desks equipped with touchscreen LCD operator interfaces, remote I/O units, digital communication infrastructure and SCADA computer(s). All components of the system are connected to a digital communications network, to provide fast and safe data exchanges. Motor drives residing on this network are under the control of PLC for a safe start and rolling operation. A multistage hot rolling process often has 10 to 30 rolling stations, and each rolling station has multiple rolls driven by motors and gear boxes. All the rolls (motors) need to be controlled in a synchronized and cascaded speed, which is designed to match the cross-section reduction ratio of the rolled bars. If a security attack occurs to alter one (or several) roll (mo- tor) speed, it may cause bar surface quality defects, stretch/compress of rolling bar before/after the speed altered station, and even severer incidents such as broken bars or cobbles in the rolling process, inducing significant production system damage or failure.

\subsection{Threat Model}
In MMS, each manufacturing machine is controlled by a Programmable Logic Controller (PLC), and those machines are interconnected in both cyber and power networks for coordinated/planned manufacturing processes. The cyber-threats may compromise the integrity of machine controllers and man- ufacturing machines through the communication networks or other means such as Trojan or insider attack. The integrity attacks are stealthy and difficult to detect in the cyber-space, but may be observed in the physical world or signals.

In our threat model, integrity attacks may be launched from Stuxnet-like malware (e.g. \cite{langner2011stuxnet}) or malicious insiders. Malware can compromise PLC programs. In addition, we assume that insiders can compromise PLC source code to intentionally result in integrity violations (e.g., PLC logic bombs \cite{govil2017ladder}). It is found in survey studies \cite{ginter2017top, perelman2016top} that insider attacks are top security challenges for air-gapped ICS (Industry Control Systems). As a result, (a) PLC source code and configurations may not be trustworthy; (b) the SCADA computer may not be trustworthy either. In our threat model, integrity attacks may also be launched from a manufacturing communication network. For example, the widely-known Maroochy Water Services Attack \cite{abrams2008malicious} injects bogus radio commands (i.e. messages) to the sewage equipment, causing 800,000 liters of raw sewage to spill out into local parks, rivers and even the grounds of a hotel. This example indicates that network intrusions should be a main part of our threat model. Regarding what are trusted, we assume that the rest of the manufacturing system, including the digital motor drives, the remote I/O units, the DSP components, the RTOS (real-time operating systems), as well as our data collection mechanisms (e.g., the sensors) are trusted.

The integrity attacks, in general, may bypass the traditional cyber-security measures but could have been detected by monitoring physical-system signals, such as electric waveform and product quality signals in MMS. To further trace down the root causes, the system needs to audit cyber signals and perform taint analysis. The root causes could include intrusion attack, malwares, trojans, insider or hardware faults (which is not a cyber-threat).

\subsection{State of the Art}
The existing cyber-security research \cite{wu2017taxonomy, wu2018cybersecurity, elhabashy2019cyber} most relies on cyber signals, integrating physical signals for cyber-threat detection is still in its infancy. Methods were developed to detect cyber threats by monitoring and analyzing structural health of the finished parts \cite{vincent2015trojan}, multiple physical signals of manufacturing process (such as acoustic signals, vibration, magnetic intensity, coal feed) \cite{wang2018sensor, gao2018watching}, vision and acoustic signals \cite{wu2019detecting}. Traditionally, quality controls are used in MMS to reduce product variability and detect the equipment aging or failure with the goal to ensure the quality of manufacturing processes. In recent years, efforts are made to use quality control methods to address cyber security issues in manufacturing system \cite{wells2014cyber, pan2017taxonomies, elhabashy2019cyber, zeltmann2016manufacturing}. While these works provide crucial insights in cyber-physical security in manufacturing, they are limited in analyzing the effect of cyber threat on a single existing quality control tool, such as control chart. As MMSs are increasingly vulnerable to cyber threats, those existing quality controls are not effective to detect the malicious cyberattacks that cause machine failure and quality degradation \cite{elhabashy2019cyber, elhabashy2019quality}. There lacks a generalized methodology to detect and diagnose cyber and physical threats in manufacturing systems.

The cybersecurity of Industrial Control Systems (ICS), including PLC security, has been drawing increasing attention in the research community (e.g., \cite{garcia2017hey, feng2019systematic, zhang2019towards}). The existing works can be classified into 6 bodies. 
\begin{itemize}
    \item Novel cyber-attacks: In \cite{garcia2017hey}, a Rootkit attack is proposed to compromise PLCs. In \cite{formby2017out}, ransomware for ICS are proposed. In \cite{urbina2016limiting}, stealthy attacks (e.g., gradually decreasing the system integrity) on ICS are proposed. In \cite{soltan2018blackiot}, three classes of new cyber-attacks are proposed to disrupt the power grid with an IoT botnet.
    \item Intrusion Detection: In \cite{formby2019temporal}, temporal execution behavior is analyzed to detect intrusions in PLCs. In \cite{cheng2017orpheus}, event-driven finite state machines are leveraged to detect data-oriented attacks on cyber-physical systems. In \cite{feng2019systematic}, a framework is proposed to generate invariants for anomaly detection in ICS. In \cite{aoudi2018truth}, a mechanism is proposed to detect stealthy attacks on ICS.
    \item Statically verifying PLC logic: In \cite{aiken1998detecting, biallas2012arcade, nellen2014cegar, biha2011formal, park2000formal}, PLC logic is statically verified in a formal manner.
    \item Detection of PLC safety violations: In \cite{janicke2015runtime, park2008plcstudio}, dynamic simulations of runtime behaviors are leveraged to detect PLC safety violations. In \cite{guo2017symbolic, mclaughlin2014trusted}, symbolic execution on PLC code is leveraged to detect PLC safety violations.
    \item Safety vetting of PLC code: In \cite{zhang2019towards}, a new program-analysis-based approach is proposed to generate event sequences that can be used to automatically discover hidden safety violations.
    \item Reverse engineering: In \cite{keliris2018icsref}, an ICS reverse engineering framework is proposed to reverse engineer PLC binaries.
\end{itemize}

As manufacturing machines and equipment are connected through power networks, cyber-threats will cause electrical signals/waveforms changes (that might include energy consumption, voltage, current, harmonics) in power networks. A method was developed to detect data integrity attacks modifying the G-code movement commands of 3D printing systems by monitoring current supplied to each electric machine \cite{moore2017power}; however, it is infeasible to monitor the current of each machine in a large-scale manufacturing systems. In smart grid, cyber security studies used waveform data from phasor measurement unit (PMU), microPMU, and smart meters \cite{tan2016survey, amini2015detecting, zhou2017partial, lu2018coupled, tian2018data, xun2018detectors}. While smart grid security studies provide necessary technical foundation, they mainly address the cyber threats affecting the functionality, stability, and cost of large-scale power networks rather than the function and precision of the devices and equipment, which are generally concerned in manufacturing systems. There are no existing works on analyzing electrical waveforms to detect threats in MMS. Researchers have obtained promising results from their preliminary works \cite{li2019detection, yang2019vulnerability} on cyber-security of electrical machines in power networks.


\section{Challenges \& insights }
\label{Over}

Information flows are intrinsic properties of an Multistage Manufacturing Systems (MMS). In computer security, a basic information flow tracking technique is \textit{dynamic taint analysis (DTA)} \cite{newsome2005dynamic}. DTA tracks taint propagation from one data variable (e.g., a buffer holding a HTTP request) to another. Taint propagation paths are typically determined by data flows and implicit flows in a computer program. And the union of all the taint propagation paths forms a \textit{taint graph}. It is clear that effective intrusion monitoring can be done via information flow tracking, and that taint graph could significantly enhance intrusion diagnosis. However, due to the following three gaps, the existing DTA techniques cannot be directly used in an MMS. (1) Without manufacturing-specific taint propagation rules, DTA cannot be implemented. Let's consider this MMS example: when a motor drive receives a distorted control signal from a compromised PLC, although the signal is a variable of the control logic program, the result (of the motor motion) is no longer a variable in any program. Restricted to cyberspace, no existing taint propagation rule can be used to propagate the taint onto the result. Hence, an MMS requires new rules to propagate taint from cyberspace to the physical world, and vice versa. (2) Since computer programs do not automatically track taint propagation, (dynamic) binary code instrumentation is widely used to let a program run extra taint tracking instructions. However, this often introduces substantial run-time overhead. Accordingly, due to strict real-time constraints, control logic programs running on PLCs are not suitable for (dynamic) binary code instrumentation. (3) Systematic DTA of an MMS has not yet been conducted in the literature.

\begin{figure*}[!t]
\centering
\centerline{\includegraphics[width=\textwidth]{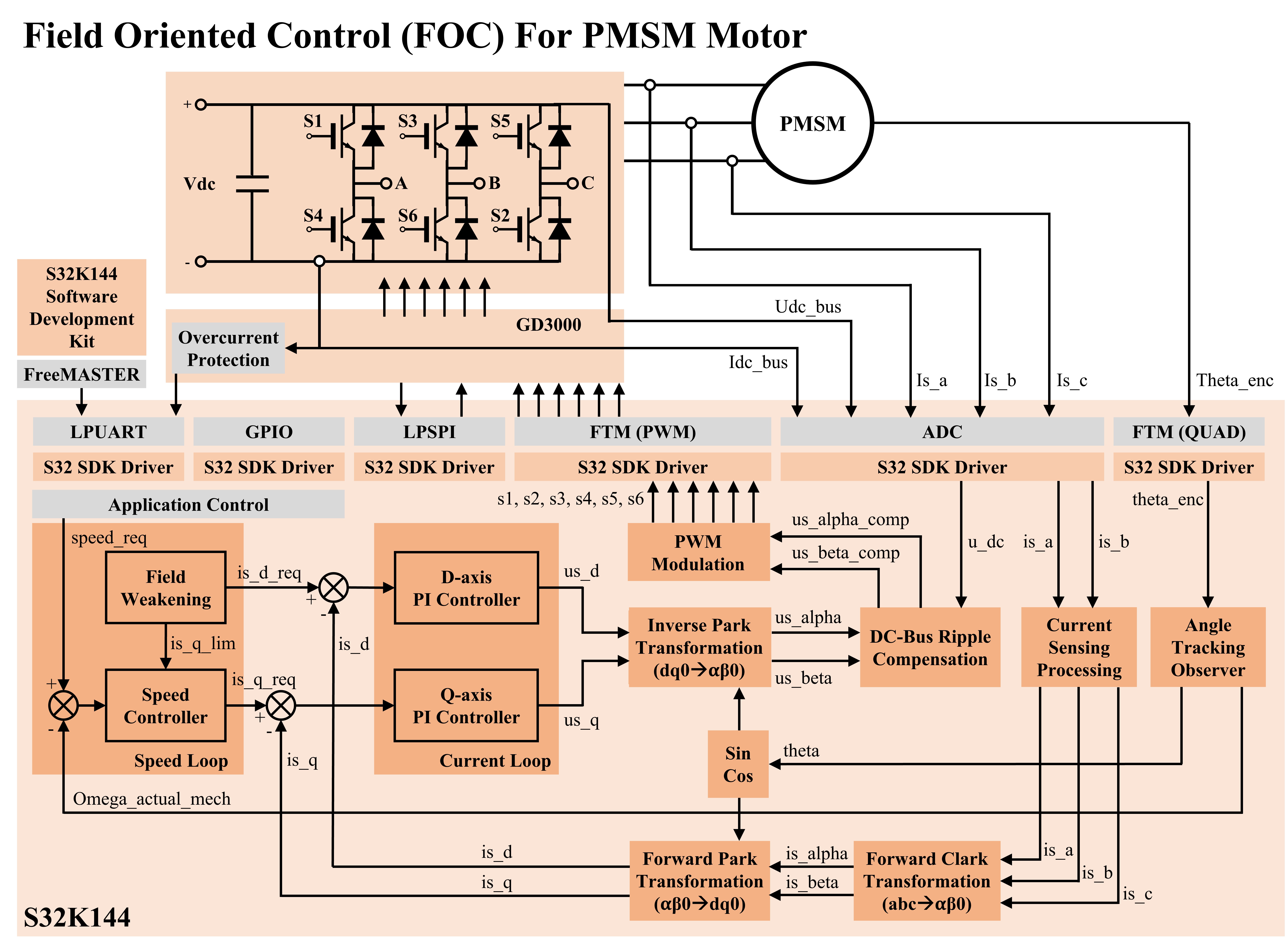}}
\caption{Control Diagram of the Mini-S\&A-Testbed}
\label{test}
\end{figure*}

There are four main intrusion diagnosis questions for an MMS to be answered: (Q1) What steps have occurred in the intrusion? (Q2) Were any other components also compromised? (Q3) What was the entry point used to gain access to the MMS? (Q4) What is the root cause? Is it due to a cyber-attack or a physical fault? In an MMS, a product being manufactured is passed through a set linear sequence of mechanical operations: these operations form a "physical production path". Since the physical path is the outcome of a rigorous engineering design process, the designer is well aware of the intended information flows along the path. However, these information flows are not modeled by existing DTA tools. In MMS, manufacturing-specific tailoring of existing forward and backward taint tracking algorithms is required. 

This brings up significant new challenges: (a) On one hand, the production path/line design literature does not provide an explicit physical-world information flow model. On the other hand, the modelling methodology of DTA is restricted to cyberspace only. (b) Even if the cyberspace taint graph were expanded to hold the information flows along the production path, the existing DTA methodology still cannot separate physical faults from cyber-threats.

The proposed model is based on two insights. (1) Some essential information flows associated with the production path are already being monitored by existing sensing components of an MMS. (2) Despite being manufacturing-specific, existing physical fault models could be leveraged to enable an information flow model to separate physical faults from cyber-threats.

\section{Mini-S\&A-Testbed \& Data Collection}
\label{UGA}

\begin{figure*}[!t]
\centering
\centerline{\includegraphics[width=\textwidth]{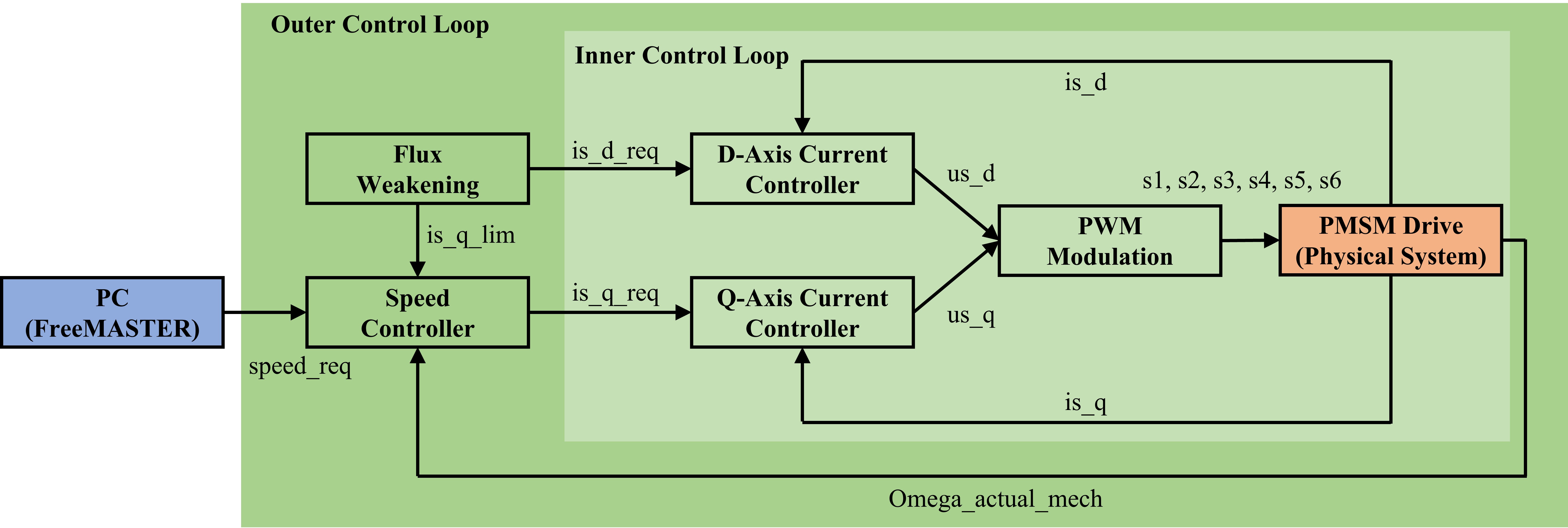}}
\caption{Control Information Flow of the Field Oriented Control for PMSM}
\label{CPS}
\end{figure*}

To test and validate the taint analysis at device level, a real-world motor drive testbed is constructed to emulate the behaviours of the industrial machines. Such testbed is called ``Mini-S\&A-Testbed" and consists of a permanent magnet synchronous machine (PMSM), a three-phase inverter and an ARM-based digital control unit. Table~\ref{tab:specification} shows the detail specifications of the testbed and Fig.~\ref{test} shows the control diagram of the motor drive.

As shown in Fig.~\ref{test}, the control unit adopts the field oriented control algorithms to regulate the rotating speed of PMSM according to the requirement from the PC. The PC and the control unit are communicating through the NXP FreeMASTER interface and the Low Power UART (LPUART) module. The control algorithms have a two-level feedback control loop, the outer control loop and the inner control loop. The outer loop has a speed regulator associated with the field weakening module to control the motor rotating speed and the air gap flux. The outer loop generates the current references and send to the inner loop. The inner loop has two proportional–integral controllers for controlling the d- and q-axis current, respectively. The outputs of the inner control loop are the d- and q-axis voltage commands. Then the Inverse Park Transformation transforms such commands into stationary reference frame, and the PWM modulation module converts these commands into PWM signals, which directly control the six power switches in the inverter. Fig.~\ref{CPS} shows a simplified diagram of the control information flow and Table~\ref{tab:variable} lists the detail descriptions of all the variables in the control algorithms.

\begin{table}[t!]
    \centering
    \caption{Mini-S\&A-Testbed Specifications}
    \begin{tabular}{c|c}
    \hline
       Control Unit  & NXP S32K144 (Arm Cortex-M4F)  \\
       \hline
       Power Module  & SMARTMOS GD3000 3-phase motor driver \\
       \hline
       Motor & LINIX 45ZWN24-40 \\
       \hline
       Motor Ratings & 24V, 40W, 4000rpm, 2.3A, 2 pole pairs \\
       \hline
       Power Supply & PS-1250APL05/S3: 12 V, 5 A \\   
       \hline
       \multirow{4}{*}{Interfaces} & On-board for CAN \\
       & On-board for LIN \\
       & On-board OpenSDA debug interface \\
       & SWD/JTAG debug interface \\
       \hline
    \end{tabular}
    \label{tab:specification}
\end{table}

In addition, the testbed has a data collection module as. Such module uses the National Instrument c-DAQ compact data acquisition device to collect and pre-process the physical measurements from the pre-deployed sensors. 

\begin{table}[t!]
    \centering
    \caption{Control Variable Detail Descriptions}
    \begin{tabular}{c|c}
    \hline
       speed$\_$req  & motor rotating speed command \\
       omega$\_$actual$\_$mech  & feedback motor rotating speed \\
       is$\_$d$\_$req  & d-axis current reference \\
       is$\_$q$\_$req  & q-axis current reference \\
       is$\_$q$\_$lim  & q-axis current limitation \\
       is$\_$d  & feedback d-axis current \\
       is$\_$q  & feedback q-axis current \\
       is$\_$a  & feedback motor phase-A current \\
       is$\_$b  & feedback motor phase-B current \\
       is$\_$c  & feedback motor phase-C current \\
       is$\_$alpha  & feedback $\alpha$-axis current \\
       is$\_$beta  & feedback $\beta$-axis current \\
       us$\_$d  & d-axis voltage command \\
       us$\_$q  & q-axis voltage command \\
       us$\_$alpha  & uncompensated $\alpha$-axis voltage command\\
       us$\_$beta  & uncompensated $\beta$-axis voltage command \\
       us$\_$alpha$\_$comp  & compensated $\alpha$-axis voltage command \\
       us$\_$beta$\_$comp  & compensated $\beta$-axis voltage command \\
       u$\_$dc  & feedback DC bus voltage \\
       theta  & feedback motor rotor position \\
       theta$\_$enc  & rotor position signal from encoder \\
    \hline
    \end{tabular}
    \label{tab:variable}
\end{table}

\section{Information Flow Model}
\label{Infor}


\begin{figure*}[!htbp]
\centering
\centerline{\includegraphics[width=0.7\textwidth]{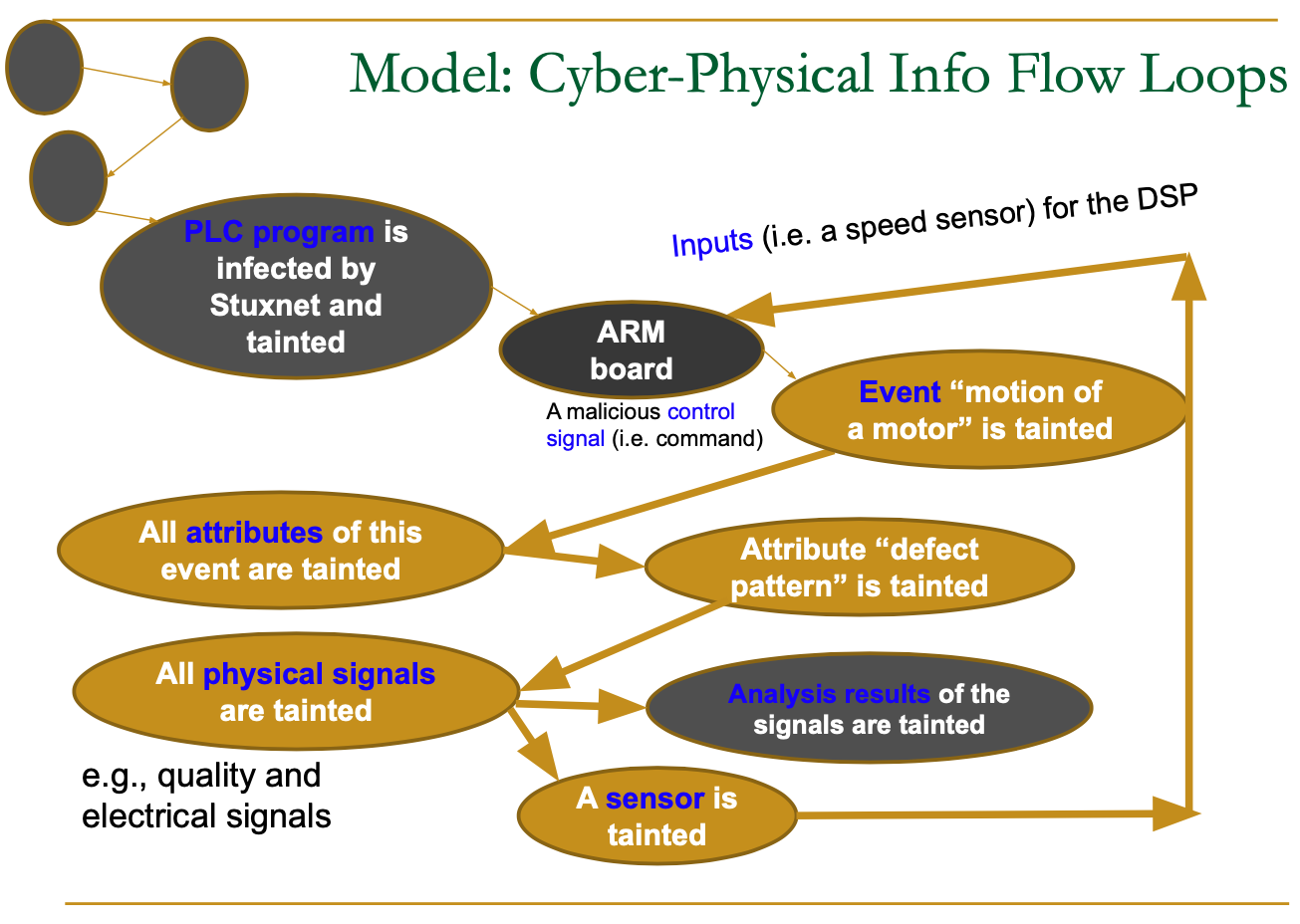}}
\caption{Cyber-Physical Info Flow Loops}
\label{infor}
\end{figure*}

This section is to introduce our proposed information flow model and how it could be applied to the min-testbed. There will be three parts: 
(1) What are the Cyber-Physical Information Flow Loops;
(2) the mapping relationship between the real testbed (Figure \ref{test}) and its control information flow (Figure \ref{CPS}); 
(3) explanations on how specific components of Figure \ref{CPS} corresponds to the five new DTA notions in our information flow model (Figure \ref{infor}) defined in Section~\ref{Infor};  

\subsection{Cyber-Physical Information Flow Loops}
As shown in Figure \ref{infor}, from the viewpoint of the MMS sensing components, the proposed information flow model is anchored by five new DTA notions: events, attributes, defect patterns, (physical) signals, and analysis results. In particular, (a) when a harmful \textit{event} (e.g., motion of a motor) is tainted, all of the event’s \textit{attributes} should be tainted. An event is recorded at every millisecond in our current data set. (b) Since one attribute of the event is usually a particular \textit{defect pattern} associated with the product, the defect pattern should be tainted. (c) When the defect pattern is being sensed, the \textit{physical signals} (e.g. quality and electrical signals) should be tainted. (d) All the \textit{analysis results} of any tainted signals should be tainted. For example, a previous event at an upstream motor is a main cause of the defect. (e) As soon as the previous event is tainted, we will reuse the above-mentioned rules to further propagate the taint. 

\subsection{Mapping Relationship}
To better understand the information flow embedded in the testbed, we have this mapping relationship as the bridge to connect the testbed and the information flow together. As shown in Figure \ref{test}, there are three main components in the current testbed setup: a PC running FreeMASTER, a S32K144 ARM board, and a motor. The PC communicates with the ARM board through LPUART (Figure \ref{test}), which corresponds to the PC provides speed reference to Speed Controller on the ARM board(Figure \ref{CPS}). The ARM board, as the cyber-space, sends the PWM control signals to the motor through the FTM(PWM) module (Figure \ref{test}), which corresponds to the current controller sends PWM control signals to winding voltage in the motor drive (Figure \ref{CPS}). The motor, as the physical world, sends the current signals and the speed signal back to the ARM board through ADC module and FTM(QUAD) module respectively, which can be mapped to winding current sends current signals back to the current controller and rotor speed sends the speed signal back to the speed controller correspondingly. In this way, we could explicitly know how the information flows through the testbed. 

\subsection{Control Information Flow}

Then, we will demonstrate how the control information flow is depicted with our proposed information flow model. Since our current focus is on how to use the collected data to generate the taint graph, we use a PC running FreeMASTER to represent the infected PLC program to send a malicious command to the ARM board as the attack scenario. Under such an assumption, we explain what the five new DTA notions in our information flow model stand for in the control information flow. Specifically, 

\begin{itemize}
    \item (Figure \ref{infor}) When a suspicious harmful event occurred (e.g., unexpected speed change of a motor), which corresponds to the change of rotor speed in Figure \ref{CPS}, and such an event will be tainted firstly.
    \item Usually an event has its intrinsic attributes like some intermediate parameters in the control testbed (e.g., Us\_d, Us\_q in Figure \ref{test}) as well as a particular defect pattern associated with the product, and the defect pattern should also be tainted.
    \item When the defect pattern is being sensed, the associated physical signals should be tainted (e.g., quality and electrical signals), which corresponds to the current signals and speed signal in Figure \ref{CPS}.
    \item All the analysis results of the tainted signals should be tainted. For example, the statistical analysis results could reveal that a previous event at an upstream motor is a main cause of the defect.
    \item (Figure \ref{infor}) As long as a sensor is tainted (e.g., current sensor and encoder in Figure \ref{CPS}), the taint will be propagated to the next event as part of the inputs.
\end{itemize}

At this point, we have acknowledged how the Cyber-Physical information flow loops express the information flow in the mini-testbed. Next, we will dive into the details about how to generate the taint graph.

\begin{figure*}[!htbp]
\centering
\centerline{\includegraphics[width=0.8\textwidth]{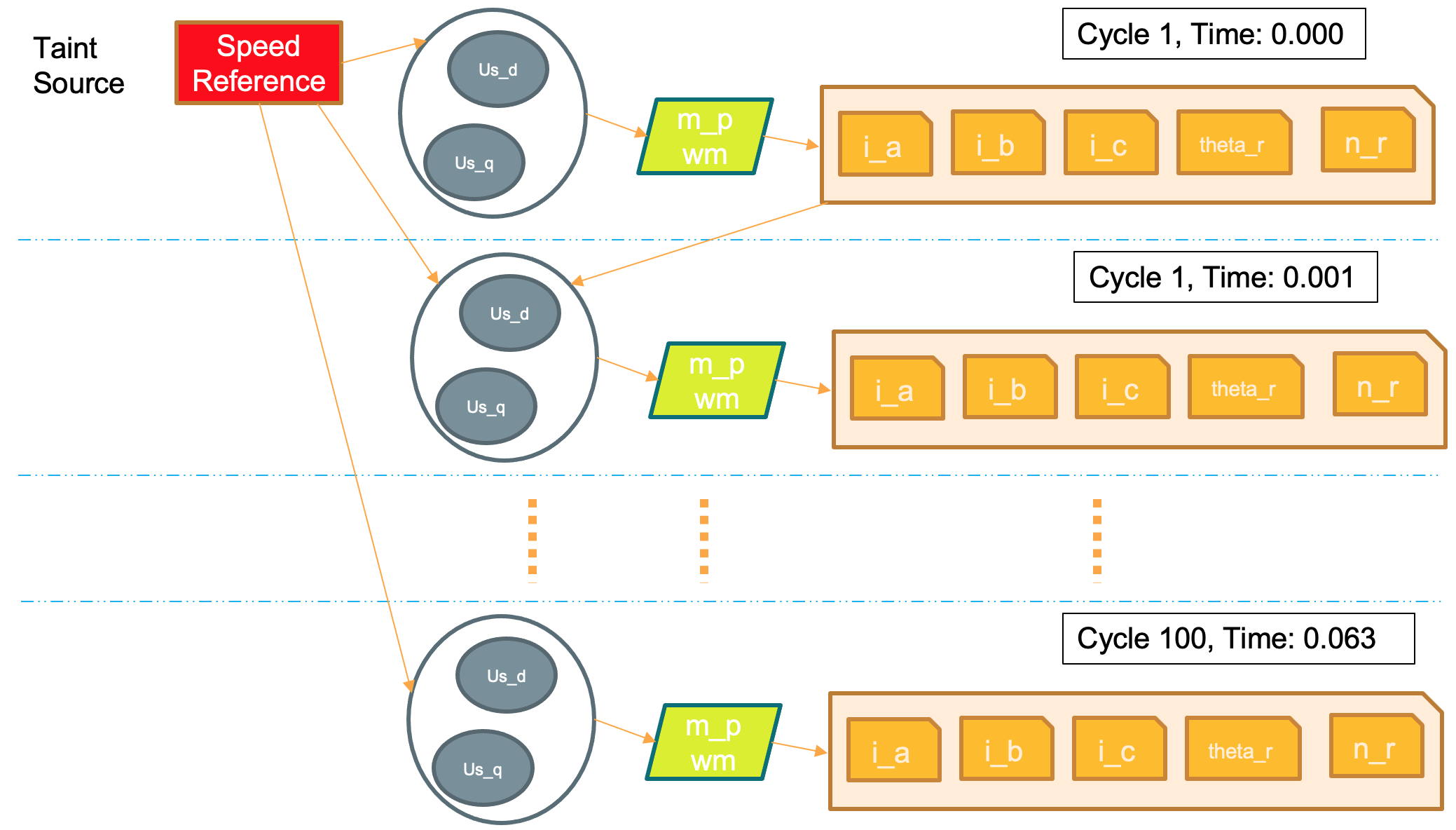}}
\caption{Sample Taint Graph}
\label{Sample}
\end{figure*}


\section{Taint Graph Generation}
\label{Taint}


The taint graph is generated based on the testbed mentioned in Section~\ref{UGA}. There will be three parts in this section: 
(1) a description of the collected dataset; 
(2) an introduction of what is taint graph; 
(3) design and implementation of our software tool on how to generate the taint graph based on the model and manufacturing-specific taint propagation rules.


\subsection{Collected Dataset}
As mentioned in Section \ref{UGA}, a set of the physical measurements has been collected from the pre-deployed sensors on the testbed. Such a dataset is used to construct our taint graph aligning with the Cyber-Physical information flow loops.
The dataset is recorded in 99 time cycles and every cycle has 64 milliseconds. In every millisecond, 8 variables are collected, which include is\_a, is\_b, is\_c, theta, omega\_actual\_mech, us\_d, us\_q, and m\_PWM. The detailed description of the first 5 variables can be found in Table \ref{tab:variable}, the last one m\_PWM represents the modulate index for PWM control. Basically, the modulate index is used to generate the switching signal (s1-s6 in the diagram), but the index is not shown in the diagram \ref{test}.

\subsection{Taint Graph Concept}
At this point, it is known that how the information flow model expresses the information flow of the testbed. As shown in Figure \ref{Sample}, basically, the taint graph is to explicitly show the information flow embedded in the testbed.
The whole taint graph is generated following the time sequence of the collected data. At every specific millisecond (marked as one event), there is one sub-graph for the information flow of the current running testbed. For every sub-graph, we have taint source node, taint propagation, and taint sink node. The followings are how specifically the whole taint graph is constructed based on the proposed Cyber-Physical information flow model and the acquired data set of the testbed.

\begin{table*}[t!]
    \centering
    \caption{Dataset Snippet}
    \begin{tabular}{c|c|c|c|c|c|c|c|c}
    \hline
       Seconds  & us\_d(2) & us\_q(3) & m\_PWM(4) & is\_a(5) & is\_b(6) & is\_c(7) & theta(8) & moega\_actual\_mech(9) \\
       \hline
       0  & 0.0591724 & -3.25714 & 0.490139 & 0.0040703 & 0.0557747 & -0.059845 & 63.3087 & 1023.32 \\
       \hline
       0.001 & 0.00780477 & -3.29621 & 0.411041 & -0.209606 & 0.162613 & 0.0469933 & 178.248 & 1001.01 \\
       \hline
       0.002 & -0.0758674 & -3.27923 & 0.341721 & -0.316444 & 0.299976 & 0.016468 & -32.333 & 961.48  \\
       \hline
       0.003 & -0.0818229 & -3.21599 & 0.290157 & -0.163818 & 0.406815 & -0.242996 & -159.07 & 954.77 \\   
       \hline
        0.004 & 0.011224 & -3.18299 & 0.273483 & -0.011192 & 0.345764 & -0.334572 & -84.919 & 996.79 \\
       \hline
       ... & ... & ... & ... & ... & ...& ... & ... & ...
    \end{tabular}
    \label{dataset}
\end{table*}

\begin{table*}[t!]
    \centering
    \caption{Graph Infor at 0.000}
    \begin{tabular}{c|c|c|c|c|c|c|c}
    \hline
       Source Node(S)  & Destination Node(D) & Value of S & Value of D & Taint Label(S) & Taint Label(D) & Node Type(S) & Node Type(D) \\
       \hline
       0  & 1 & N/A & 1.0 & 1 & 0 & ts(taint source) & arm   \\
       \hline
        1 & 2 & 1.0 & 0.0591724 & 0 & 0 & arm & an(attribute node) \\
       \hline
       1 & 3 & 1.0 & -3.25714 & 0 & 0 & arm & an  \\
       \hline
        2 & 4 & 0.0591724 & 0.490139 & 0 & 0 & an & cs(control signal) \\   
       \hline
       3 & 4 & -3.25714 & 0.490139 & 0 & 0 & an & cs \\
       \hline
        ... & ... & ... & ... & ... & ...& ... & ...\\
        \hline
        5 & 1 & 0.0040703 & 1.0 & 0 & 0 & sd(sensor data) & arm\\
        \hline
        6 & 1 & 0.0557747 & 1.0 & 0 & 0 & sd & arm\\
        \hline
        ... & ... & ... & ... & ... & ...& ... & ...\\
        \hline
        9 & 1 & 1023.32 & 1.0 & 0 & 0 & sd & arm\\
    \end{tabular}
    \label{input}
\end{table*}

\subsubsection{Taint Source Selection}
In the existing DTA techniques \cite{newsome2005dynamic}, taint source refers to the one where the untrusted data or malicious command is introduced. In our information flow, since the received malicious command can modify different parameters to launch different attacks. For example, in Figure \ref{test}, if speed reference is modified, which definitely will cause a series of changes in the control loop, and this speed reference will be marked as taint source in this attack scenario. If current reference is modified, the control loop will have the corresponding changes, and this current reference will be marked as taint source in this attack scenario. In a word, taint source selections mean different attack scenarios and generate different taint graphs.

\subsubsection{Taint Sink Node}
In the sub-graph for every specific millisecond, there are taint sink nodes where the information flows to. In our proposed information flow model, the sensor data (e.g., current sensor, speed sensor in Figure \ref{CPS}) will be marked as taint sink nodes.

\subsubsection{Other Node}
Besides the taint source node and taint sink node, there are two more types of node, one is PWM control signal \textit{m\_pwm}, which is sent from the ARM board to the motor drive, the other one is the intermediate parameters {\color {red} named as attribute node} in the control loop (e.g., Us\_d, Us\_q), which are also regarded as attributes of one specific event.

\subsubsection{Hierarchy Node System}
There is a hierarchy node system for the whole taint graph for the convenience of retrieving values of the specific nodes at the specific millisecond: a) The data is collected in the unit of cycle as the testbed runs; b) Every cycle has sixty-four milliseconds, and each millisecond is considered as one event; c) In addition, each node in the sub-graph has its variable ID. In a word, each node in the whole taint graph has a unique Node ID consisting of cycle ID (prefix), event ID (time stamp), Variable ID (suffix). For example. in cycle 66, and $35^{th}$ millisecond, node \textit{m\_pwm} has Node ID of 66-35-04.

\begin{figure*}[htbp]
\centering
\centerline{\includegraphics[width=0.8\textwidth]{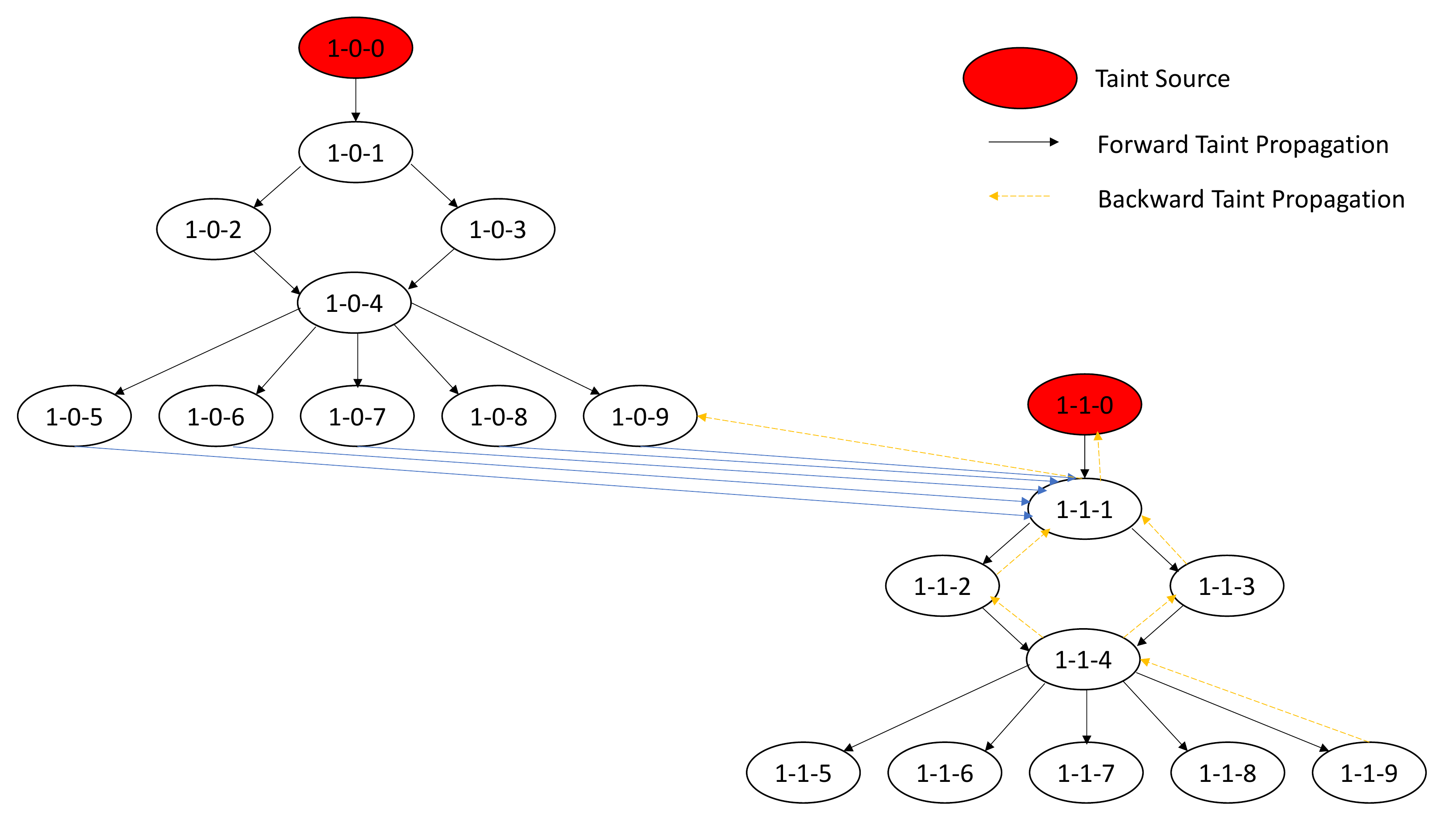}}
\caption{Two Sub-graphs with connection}
\label{taintgraph}
\end{figure*}

\subsection{Design and Implementation of the Software Tool}
To construct the taint graph, we developed a software tool written in Python. There are three stages to generate the {\color{blue} taint graph}: 1) reorganize the dataset and assign the variable IDs to the collected 8 variables; 2) prepare a file for the event at 0 millisecond including the connection information between these variables, both taint information and node type of these variables, and use this file as input into our reading procedure to generate the same kind of file, which contain the above information for all recorded events; 3) the generated file will be fed into our tainting procedure to construct the taint graph.

\subsubsection{Preliminary process of the dataset}
For our testbed, This stage is to assign the variable IDs to the collected 8 variables, e.g., 0 for speed reference, 1 for the ARM board, 2-9 for us\_d, us\_q, m\_PWM, is\_a, is\_b, is\_c, theta, and omega\_actual\_mech, as shown in Table \ref{dataset}. 

\subsubsection{Input Preparation}
As shown in Table \ref{input}, prepare a file for the event at 0 millisecond including the connection information between these variables, both taint information and node type of these variables, and use this file as input into our reading procedure to generate the same kind of file, which contain the above information for all recorded events.

\subsubsection{Taint Graph Generation}
At moment of zero second of cycle 1, the first sub-graph starts with our selected taint source (speed reference). As mentioned before, the whole taint graph is generated following the time sequence of the collected data. The single sub-graph will be connected to its adjacent one following the time sequence. The whole taint graph is formed until the end of the last sub-graph in our data set.

\paragraph{Sub-graph} For every event (at the specific millisecond), there will be one corresponding sub-graph. Notably, There are two exceptions of the sub-graph construction. One is the first sub-graph, which only has taint source as its input, while the following sub-graphs all have both taint source and sensor data (taint sink) as their inputs. The other one is that the last sub-graph does not have out edges to connect the next sub-graph. As shown in Figure \ref{Sample}, the first sub-graph starts with our selected taint source and connects to the attribute nodes followed by control signal node, which connects to sensor data node. 

\paragraph{The Whole Taint Graph} Based on the control information flow in Figure \ref{CPS}, both taint source and sensor data will be used as inputs for the next event to calculate a new set of attribute node, control signal node, and sensor data (sink node). The whole taint graph is finalized after the last event.



\section{Cause Analysis}
\label{Case}
In this section, we will explain how to utilize the generated taint graph to do preliminary cause analysis through forward and backward taint tracking. There are two types of intrusion diagnosis questions to be answered, which are (a) ``if X is compromised, would Y be part of the damage?" (b) ``if damage Y is observed, would X be a causing factor?". Case study I is to demonstrate how forward taint tracking identify the damages caused by X, and case study II explains how to decide whether X should account for damage Y or not through backward taint tracking.

\subsection{Case Study I}
When a malicious command (e.g., modifying the speed reference to a bad one) is launched at the ARM board, through forward taint tracking on the sample taint graph, a series of damage and compromised components could be identified. 

To answer the question ``if the speed reference is maliciously modified, what is the damage?", we have the following answers.
\begin{itemize}
    \item In Figure \ref{taintgraph}, once the speed reference node (1-0-0) is maliciously modified, following the forward taint propagation, the arm boar node (d 1-0-1), the connected attribute node (1-0-2) \& (1-0-3), \textit{m\_PWM} node, the connected sensor data node (1-0-5), (1-0-6), (1-0-7), (1-0-8), (1-0-9) will all be part of the damage.
    \item Furthermore, because the mini-testbed utilizes the control information flow in Figure \ref{CPS}, so both the maliciously modified speed reference and the generated sensor data will serve as the inputs for the next event, which means all the following events will all belong to the damage. 
\end{itemize}


\subsection{Case Study II}
However, in the real world setting, some attacks can not be detected in real-time, which means some damage is observed firstly, and then intrusion diagnosis steps in to identify how the intrusion happened and what the root cause is. 
For example, if defect Y was found in one product produced by motor M at time T, how could we identify the causing factors and even the root cause?

To answer the question ``if damage Y is observed, would X be a causing factor?", we have the following answers. 
\begin{itemize}
    \item In our collected dataset, variable omega\_actual\_mech is the feedback motor rotating speed, which directly impacts the product quality. In Figure \ref{taintgraph}, once we know at what time the damage Y is observed, we could locate the abnormal sensor data node omega\_actual\_mech (aa-bb-09), which will be used as the taint source node for backward taint tracking.
    \item Through backward taint tracking on the taint graph, as shown in Figure \ref{taintgraph}, the taint starting from the abnormal sensor data node omega\_actual\_mech could be firstly propagated from (1-1-9) to (1-1-4). Through (1-1-2) \& (1-1-3), the taint is propagated from (1-1-4) to (1-1-1). At this point, the taint will be propagated within the current sub-graph to (1-1-0) and to the sensor data nodes in the previous event respectively. In the former case, the speed reference could be one the candidates who account for the damage Y.
    \item For the latter case, the taint will be propagated through the events backwardly. During taint propagation, the values of those tainted nodes, e.g., (1-0-4), (1-0-5), (1-0-7), are compared with their normal values recorded under the same testbed configuration. Some obvious mismatching will be marked as suspicious candidates causing the damage Y. 
\end{itemize} 


\section{Conclusions}
\label{Con}

In this work, we conduct a case study which (a) extends the existing DTA method with manufacturing-specific taint propagation rules, and (b) applies the extended method to perform preliminary intrusion diagnosis with a small-scale test-bed. 
Using the customized set of manufacturing-specific taint propagation rules, we have developed a software tool to construct a taint graph directly based on 
the event data collected from the testbed. 
Through answering several specific intrusion diagnosis questions, including (a) ``if control signal X is compromised, would Y be part of the damage?" and (b) ``if damage Y is observed, would X be a causing factor?", we have partially evaluated the soundness and effectiveness of the proposed cyber-physical DTA method through a case study.

\bibliographystyle{ieeetr}
\bibliography{IEEEexample}

\vspace{12pt}

\end{document}